\documentclass[prb,aps,floats,showpacs,amsmath,amssymb,preprint]{revtex4}
\usepackage{bm,graphicx,epsf}
\usepackage{bm,graphicx}
\newcommand{\be}{\begin{equation}}
\newcommand{\ee}{\end{equation}}   
\newcommand{\bea}{\begin{eqnarray}}
\newcommand{\eea}{\end{eqnarray}}
\newcommand{\phrl}[1]{Phys.~Rev.~Lett. {\bf #1}}
\newcommand{\phrb}[1]{Phys.~Rev.~B {\bf #1}}

\newcommand{\bib}{\bibitem}
\newcommand{\lb}{\left[}
\newcommand{\rb}{\right]}
\newcommand{\lp}{\left(}
\newcommand{\rp}{\right)}

\renewcommand{\k}{{\bf k}}

\newcommand{\E}{\bf E}

\begin{document}
\title{Effect of helical edge states on the tunneling conductance in ferromagnet/noncentrosymmetric superconductors junctions}
\author{Soumya Prasad Mukherjee }
\affiliation{Department of Theoretical Physics \\ Indian Association for the Cultivation of Science, Jadavpur, Kolkata 700 032, India}

\date{\today}


\begin{abstract}
Helical edge states exist in the mixed spin-singlet and spin-triplet phase of a noncentrosymmetric (NCS) superconductor. In this article we have considered a planar ferromagnetic metal/NCS superconductor tunnel junction and have studied the effect of these helical edge states which manifests itself through the charge and spin tunneling conductance across the junction. We have shown the behavior of conductances for the entire range of variation of $\gamma = \Delta_-/\Delta_+$ where $\Delta_\pm$ are the order parameters in the positive and negative helicity bands of the NCS superconductor.There exists a competition between the Rashba parameter $\alpha$ and the exchange energy $E_{ex}$ which is crucial for determining the variation of the conductance with the applied bias voltage across the junction. We have found a nonzero spin current across the junction which appears due to the exchange energy in the Ferromagnet and modulates with the bias voltage. It also changes its profile when the strength of the exchange energy is varied.
 
\end{abstract}

\maketitle
\section{Introduction}

Study of the tunneling conductances between different materials coupled together is an interesting area of research in condensed matter physics which is also helpful in predicting many properties of the systems which are coupled. Ferromagnetism and superconductivity show antagonistic behavior yet when they are coupled together they show some interesting transport properties. In the ferromagnetic system the presence of exchange energy creates a population imbalance between the two spin sub-bands. When a ferromagnetic metal is coupled with a normal s-wave superconductor then there occur some interesting phenomenon which were absent if the ferromagnet was replaced by the normal metal. In normal metal s-wave superconductor junction for a bias voltage less than the superconducting order parameter there occurs a process which is called Andreev reflection. By this process the incident electron with spin $\sigma$ is reflected back as a hole with same energy but with opposite spin $\overline{\sigma}$. The change in momentum due to this process is negligible as the Fermi energy in normal metal is huge. In case of ferromagnetic metal the presence of exchange energy induces an extra momentum change of the reflected hole which causes a reduction in the Andreev reflection amplitude. In other words the presence of spin sub-bands reduces the availability of the conducting channel and this reduces the Andreev reflection \cite{Beenakker,Wang}. Later this idea has been extended to the case of unconventional superconductor/Ferromagnet junction and similar observation of the suppression of the Andreev reflection was found through the appearance of dip in the zero biased differential conductance \cite{Vasko}. This fact has also been utilized to determine not only the polarization of the ferromagnetic material but also the nature of the order parameters of various kind of superconductors. It has been studied to determine the order parameter symmetry of the singlet d-wave superconductor \cite{Goldman,Dong,Venkatesan,Zhao,Igor,Valls},triplet p-wave chiral superconductor \cite{ Golubov,Manske, Sigrist,Iniotakis,Brydon,Takehashi, Gentile,Brydon1}, mixed singlet-triplet order parameter superconductor \cite{Samokhin} and recently also in superconducting Graphene system \cite{Moghaddam,Sudbo}. One peculiar feature of this Andreev reflection is that it can produce singlet as well as triplet Cooper pairs in the superconducting regions provided the superconductor in the opposite side can sustain both the singlet as well as triplet order parameters. As we will explain shortly the NCS superconductors appear as an ideal candidate here since the order parameter in them is a combination of singlet as well as triplet order parameters. This kind of study has been performed by Samokhin {\it {et.al}} \cite{Samokhin} already by calculating the bulk conductance across a ferromagnetic metal/NCS superconductor junction.

The discovery \cite{Fujimoto} of edge states with definite topological properties in the noncentrosymmetric superconductors have intensified the research in this field. The main feature of this class of materials is that in them the inversion symmetry is broken. This broken inversion symmetry gives rise to some special kind of spin-orbit coupling which is known as Rashba spin-orbit coupling(RSOC).The presence of RSOC breaks the degenerate electron band and gives rise to two bands with opposite helicities. As a result of this the superconducting order parameter instead of having any definite symmetry is a combination of singlet and triplet components. This particular combination gives rise to many strange behaviors in many of the physical properties of the system. The presence of the Rashba parameter also manifests itself by forming the topological edge states. The topological edge states exists provided the amplitude of the triplet order parameter is greater than that of the singlet order parameter.\cite{Nagaosa}. The topological edge states are itself interesting and has profound effect on the tunneling conductance\cite{Nagaosa,Mukherjee}. Many work has been done on them and there exists a spin current across the junction of a normal metal/NCS superconductor due to these topological edge states, which appears by the application of a magnetic field external to the sample. 

Another more interesting study will be to study the effect of these edge states, present in NCS superconductor, across the ferromagnet/NCS superconductor hetero-structure. There already exist works on the tunneling properties of these edge states in NCS superconductor. \cite{Eschrig,Hayashi,Nagaosa,Mukherjee}.In this article we study the effect of the helical edge states in the charge and spin tunneling conductances across a ferromagnetic metal and NCS superconductor junction. The NCS superconductor is in its mixed singlet-triplet phase. We have shown the variation of the charge and spin conductances with the applied bias and with respect to various parameters which are Rashba parameter $\alpha$, exchange energy $E_{ex}$, the ratio between the strength of the order parameters of the negative and positive helicity bands which we call $\gamma$ and the barrier height Z. We have found a number of interesting features in our study. As there exists a competition between the Rashba parameter and the exchange energy so the conductances change accordingly. Another interesting feature is the existence of a tunneling spin current across the junction which appears due to the presence of the exchange energy which breaks $g_c(\phi)\neq g_c(-\phi)$ symmetry (for angle resolved charge conductance) and $g_s(\phi)\neq -g_s(-\phi)$ (for angle resolved spin conductance) by creating an probability imbalance between the number of up-spin and down-spin incident electrons. This spin current although smaller than the magnitude of the corresponding charge current yet its value can be increased considerably by coupling a ferromagnetic material with higher value of the exchange energy. Not only that this spin current or the spin conductance modulates with the variation of the bias voltage across the junction of the ferromagnet/NCS superconductor. The direction of this spin current is reversed by creating a reverse population density in the spin sub-bands. The effect of the parameters $\alpha$ and Z are also studied qualitatively and the results are explained.

This article is organized as follows. In section II we have derived all the formula and described the method of our working. In section III we have found numerically the tunneling charge and spin conductances from a Ferromagnetic metal to a NCS superconductor due to the edge states in the NCS superconductor. We have shown here our results in the form of various diagrams and also explained them. We summarize our results in section IV.

\section{tunneling charge and spin conductances}
 We begin with the Hamiltonian for an NCS superconductor in which Cooper pairs form between the electrons 
 within the same spin-split band: 
\be
{\cal H} = \sum_{\k , \lambda = \pm} \lb \xi_{\k\lambda }c_{\k \lambda}^\dagger
c_{\k \lambda} +\lp \Delta_{\k\lambda} 
c_{\k \lambda}^\dagger c_{-\k \lambda}^\dagger + \, \rm{h. c.}
 \rp \rb  \, ,
\label{H_band}
\ee
where $\xi_{\k\lambda} = \xi_{\k}+\lambda \alpha \vert \k\vert$ for Rashba SOI \cite{Rashba}, 
$\xi_{\k} = \hbar^2\k^2/(2m)-\mu$. Here $\mu$, $m$, $\lambda$, $\k$, $\alpha$, and $\Delta_{\k\lambda}$ 
denote chemical potential, mass of an electron, spin-split band index $(\pm)$, momentum of an electron,
coupling constant of Rashba SOI given by $\hat{V}_{so}=\alpha \bm{\eta}_\k \cdot \hat{\bm{\sigma}}$ with
$\bm{\eta}_\k = \hat{\bm{y}}k_x - \hat{\bm{x}}k_y$ and the Pauli matrices $\bm{\sigma}$, 
and pair potential in band $\lambda$ respectively.
We choose $k_y +ik_x$-wave pair in both the bands, i.e., $\Delta_{\k\lambda} = \Delta_\lambda \Lambda_\k$
with $\Lambda_\k = -i\exp [-i\phi_\k]$. This corresponds to triplet component of pair potential
$\hat{\Delta}_T = (\bm{d}_\k \cdot \bm{\sigma})i\sigma_y$ with 
$\bm{d}_\k = \frac{1}{2\vert \k\vert}(\Delta_+ + \Delta_-)\bm{\eta}_\k$, i.e., the amplitude of the
triplet component $\Delta_t= \frac{1}{2}(\Delta_+ +\Delta_-)$ and the singlet component of the pair potential
is $\hat{\Delta}_S = i\Delta_s \sigma_y$ with amplitude $\Delta_s = \frac{1}{2}(\Delta_+ -\Delta_-)$ 
(Ref.\onlinecite{Mandal}). 
Therefore the superconductor is purely triplet with $k_y +i k_x$-wave
symmetry when $\Delta_+ = \Delta_-$, purely singlet with $s$-wave symmetry when $\Delta_- = -\Delta_+$,
and triplet and singlet components with equal amplitude when $\Delta_- = 0$.
Therefore the Hamiltonian (\ref{H_band}) in the matrix form \cite{Koga} read as
\be
  H = \left( \begin{array}{cc}
       \hat{h}_\k & \hat{\Delta}_\k \\
      -\hat{\Delta}^*_{-\k} & -\hat{h}^*_{-\k} \end{array} \right) \, ,
\label{H_super}
\ee
where $\hat{h}_\k = \xi_\k + \hat{V}_{so}$ and $\hat{\Delta}_\k = \hat{\Delta}_T +\hat{\Delta}_S$.
The solution of the Hamiltonian (\ref{H_super}) in the bulk 
is with the energy eigenvalues values $\pm \sqrt{\xi_+^2 + \Delta_+^2}$ and $\pm \sqrt{\xi_-^2 + \Delta_-^2}$,
in  with the Cooper pairing between electrons within the same spin-split band.
Correspondingly, there are two Fermi surfaces with Fermi momenta $k_F^\pm = \mp m\alpha /\hbar^2 +
\sqrt{(m\alpha/\hbar^2)^2 + 2m\mu/\hbar^2}$, i.e., $k_F^+ < k_F^-$.

Now we consider a planar junction between a ferromagnetic metal($x<0$) and NCS superconductor($x>0$) in such a way that the barrier is along Y direction(x=0). The interface is characterized by a potential $U(r)= U_0\delta(x)$ with $U_0$ as the strength of the barrier.The Hamiltonian describing the ferromagnet is given as 
\be
H_{FM}=(-\frac{\bigtriangledown^2}{2m}+U(r)-E_fm)\hat{\sigma_0}-\sigma h_0
\ee
where $\sigma=\pm$ for different spin orientations. This second term in the Ferromagnetic Hamiltonian is called exchange energy interaction and play a important role in the tunneling process. Here $E_{fm}$ describe the Fermi energy in the Ferromagnetic side. We neglect any Fermi energy mismatch between two sides of the junction ie. assume that the Fermi energy on both sides of the junction are at the same level. We now try to construct the quasi-particle wave functions in the Ferromagnetic side. To do this we follow the way of writing the wave function as described by \cite{Samokhin} Samokhin {\it{et al}}. We make further simplification to them by taking $E=0$ such that the wave vector for particles and holes with any particular $\sigma$ are identical. We describe the quasi-particle wave vectors for $\sigma=\uparrow,\downarrow$ as,
\bea
k^{e}_\uparrow=\sqrt{2m[E_{fm}+h_0]} \\ 
k^{e}_\downarrow=\sqrt{2m[E_{fm}-h_0]}
\eea

An incoming spin incident from the FM side on the interface may be normally reflected or Andreev reflected. If we denote by $r^{\sigma \uparrow}_e$ and $r^{\sigma \downarrow}_e$ the normal reflection coefficients , by $r^{\sigma \uparrow}_h$ and  $r^{\sigma \uparrow}_h$ the Andreev reflection coefficients for an incoming spin $\sigma$ then we can write the spin-up and spin-down quasi-particle wave functions as ,

\bea
\Psi_{FM}^{\uparrow}(x) = \left( \begin{array}{c} 1 \\  0 \\  0 \\ 0 \end{array} \right) \ e^{k_{\uparrow}^e cos\theta x}+r^{\uparrow \uparrow}_e\left( \begin{array}{c} 1 \\  0 \\  0 \\ 0 \end{array} \right) \ e^{-k^{e}_\uparrow cos\theta x}+r^{\uparrow \downarrow}_e\left( \begin{array}{c} 0 \\  1 \\  0 \\ 0 \end{array} \right) \ e^{-k^{e}_\downarrow cos\theta_{\downarrow}^e x}\nonumber\\+r^{\uparrow \uparrow}_h\left( \begin{array}{c} 0 \\  0 \\  1 \\ 0 \end{array} \right) \ e^{k^{e}_\uparrow cos\theta_{\uparrow}^e x}+r^{\uparrow \downarrow}_h\left( \begin{array}{c} 0 \\  0 \\  0 \\ 1 \end{array} \right) \ e^{k^{e}_\downarrow cos\theta_{\downarrow}^e x}
\eea

where we have kept room for all possible spin flip processes at the interface. Here $\theta^{e}_\uparrow$ and $\theta^{e}_\downarrow$ are respectively the angles made by the wave-vectors $k_{\uparrow}^e$ and $k_{\downarrow}^e$ respectively with the interface normal. Similarly the wave function for spin-down quasi-particle is written as,

\bea
\Psi_{FM}^{\downarrow}(x) = \left( \begin{array}{c} 0 \\  1 \\  0 \\ 0 \end{array} \right) \ e^{k_{\downarrow}^e cos\theta x}+r^{\downarrow\downarrow}_e\left( \begin{array}{c} 0 \\  1 \\  0 \\ 0 \end{array} \right) \ e^{-k^{e}_\downarrow cos\theta x}+r^{\downarrow\uparrow}_e\left( \begin{array}{c} 1 \\  0 \\  0 \\ 0 \end{array} \right) \ e^{-k^{e}_\uparrow cos\theta_{\uparrow}^e x}\nonumber\\+r^{\downarrow\downarrow}_h\left( \begin{array}{c} 0 \\  0 \\  0 \\ 1 \end{array} \right) \ e^{k^{e}_\downarrow cos\theta_{\downarrow}^e x}+r^{\downarrow\uparrow}_h\left( \begin{array}{c} 0 \\  0 \\  1 \\ 0 \end{array} \right) \ e^{k^{e}_\uparrow cos\theta_{\uparrow}^e x}
\eea

 We can also write the quasi-particle wave functions in the NCS superconductor side which describe the edge states. They are 

\bea
\Psi_S (x,y) &=& e^{ik_yy} [ e^{-\kappa_+x} \{ c_1\, \psi_e^+ e^{ik^+_{Fx}x} + c_2\, \psi_h^+ e^{-ik^+_{Fx}x} \}
  \nonumber\\ & & + e^{-\kappa_-x} \{ d_1 \,\psi_e^- e^{ik^-_{Fx}x} + d_2\, \psi_h^- e^{-ik^-_{Fx}x} \} ] \, ,
\eea

where Fermi momenta along $x$-direction in two spin-split bands are $k_{Fx}^\pm = \sqrt{k_F^{\pm 2}-k_y^2}$. Quasiparticle and quasihole wave functions \cite{Mukherjee} in two spin-split bands $(\pm)$ are given by

\bea
 & &\psi_e^+ = \left( \begin{array}{c} u_+ \\  - ie^{i\phi_{+\sigma}}u_+ \\  ie^{i\phi_{+\sigma}}v_+ \\ v_+ \end{array} \right) \, ,\,
 \psi_h^+ = \left( \begin{array}{c} v_+ \\  + ie^{-i\phi_{+\sigma}}v_+ \\ - ie^{-i\phi_{+\sigma}}u_+ \\ u_+ \end{array} \right) \, ,
\label{psi_plus} \\
 & &\psi_e^- = \left( \begin{array}{c} u_- \\   ie^{i\phi_{-\sigma}}u_- \\  ie^{i\phi_{-\sigma}}v_- \\ -v_- \end{array} \right) \, ,\,
 \psi_h^- = \left( \begin{array}{c} v_- \\  - ie^{-i\phi_{-\sigma}}v_- \\  -ie^{-i\phi_{-\sigma}}u_- \\ -u_-\end{array} \right) \, ,
\label{psi_minus}
\eea

where $\frac{u_+}{v_+} = (E-i\Gamma_+)/\Delta_+$, $\frac{u_-}{v_-} = (E-i\Gamma_-)/\Delta_-$, and $\Gamma_\pm = \sqrt{\Delta_\pm^2-E^2}$
for an edge state with energy $E$, and $\sin (\phi_{\pm\sigma}) = k_{\sigma}^e/k_F^\pm$, where $\sigma$ stands for the spins of the incoming electron. Here $c_1$, $c_2$, $d_1$, and $d_2$ are the corresponding weights at which these four quasi-particle and quasihole states mix, and $\kappa_\pm = m\Gamma_\pm /k_{Fx}^\pm$  are the inverse of the length scales of localized edge state for two spin-split bands.

We have to determine the above coefficients by matching the boundary conditions suitably. In doing this we have to consider the spin of the incoming electrons.The reflection amplitudes can be found out by matching the wave functions and the velocity flux at $x=0$:

\be
\Psi_{FM}^\sigma(x=0,y) = \Psi_S(x=0,y)  \,  , \label{BC1} 
\ee

\bea
 \left( \begin{array}{cccc}
-\frac{i}{m}\partial_x & 0 & 0 & 0 \\
0 & -\frac{i}{m}\partial_x & 0 & 0 \\
0 & 0 & \frac{i}{m}\partial_x & 0 \\
0 & 0 & 0 & \frac{i}{m}\partial_x 
\end{array} \right) \Psi_{FM}^\sigma (x,y)\vert_{x=0}
=2iU \left( \begin{array}{rrrr}
1 & 0 & 0 & 0 \\
0 & 1 & 0 & 0 \\
0 & 0 & -1 & 0 \\
0 & 0 & 0 & -1 
\end{array} \right) \Psi_{FM}^\sigma(x=0,y)\nonumber \\
 + \left( \begin{array}{cccc} 
-\frac{i}{m}\partial_x & i\alpha & -i\frac{\Delta_t}{k_F} & 0 \\
-i\alpha & -\frac{i}{m}\partial_x & 0 & -i\frac{\Delta_t}{k_F} \\ 
i\frac{\Delta_t}{k_F} & 0 & \frac{i}{m}\partial_x & -i\alpha \\
0  & i\frac{\Delta_t}{k_F} & i\alpha & \frac{i}{m}\partial_x 
\end{array} \right) \Psi_S (x,y)\vert_{x=0} \,\, . \label{BC2}
\eea

From the conservation of the parallel component of the wave vector one can write for an incoming up-spin particle,
\be
k_{\uparrow}^e sin\theta = k_{\downarrow}^e sin\theta_{\downarrow}^e=k_{\uparrow}^e sin\theta_{\uparrow}^e=k_{F+} sin\theta_{+\uparrow} = k_{F-} sin\theta_{-\uparrow}
\ee


Similarly for an incoming down-spin particle we have,
\be
k_{\downarrow}^e sin\theta = k_{\uparrow}^e sin\theta_{\uparrow}^e=k_{\downarrow}^e sin\theta_{\downarrow}^e=k_{F+} sin\theta_{+\downarrow} = k_{F-} sin\theta_{-\downarrow}
\ee


This is a peculiarity that appears in the case of a Ferromagnetic metal is that the presence of exchange energy causes an imbalance between the density of states of the up and down spin electrons. As a result the probability that an incoming electron is spin up($P_\uparrow$) differs from that of an electron with spin down ($P_\downarrow$). This probability factor is given as $P_\sigma=\frac{1}{2}(1+\sigma h_0/E_{FM})$.

Extending the generalized BTK formalism\cite{BTK} in our case we can write the angle resolved charge($G_{c\uparrow}(E,\theta)$) and the spin($G_{s\uparrow}(E,\theta)$) conductance as(for an $\uparrow$ electron incoming)
\be
G_{c\uparrow}(E,\theta)=1+(|r_{h}^{\uparrow\uparrow}|^2-|r_{e}^{\uparrow\uparrow}|^2)+(\frac{tan\theta}{tan\theta_{\downarrow}^e})(|r_{h}^{\uparrow\downarrow}|^2-|r_{e}^{\uparrow\downarrow}|^2)
\ee

\be
G_{s\uparrow}(E,\theta)=(|r_{h}^{\uparrow\uparrow}|^2-|r_{e}^{\uparrow\uparrow}|^2)+(\frac{tan\theta}{tan\theta_{\downarrow}^e})(|r_{e}^{\uparrow\downarrow}|^2-|r_{h}^{\uparrow\downarrow}|^2)
\ee

Similarly the angle resolved charge($G_{c\downarrow}(E,\theta)$) and the spin($G_{s\downarrow}(E,\theta)$) conductance as(for an $\downarrow$ electron incoming)
\be
G_{c\downarrow}(E,\theta)=1+(|r_{h}^{\downarrow\downarrow}|^2-|r_{e}^{\downarrow\downarrow}|^2)+(\frac{tan\theta}{tan\theta_{\uparrow}^e})(|r_{h}^{\downarrow\uparrow}|^2-|r_{e}^{\downarrow\uparrow}|^2)
\ee

\be
G_{s\downarrow}(E,\theta)=(|r_{h}^{\downarrow\downarrow}|^2-|r_{e}^{\downarrow\downarrow}|^2)+(\frac{tan\theta}{tan\theta_{\uparrow}^e})(|r_{e}^{\downarrow\uparrow}|^2-|r_{h}^{\downarrow\uparrow}|^2)
\ee

Here the factor $\frac{tan\theta}{tan\theta_{\downarrow}^e}$ for $\uparrow$-spin case and $\frac{tan\theta}{tan\theta_{\uparrow}^e}$ for $\downarrow$-spin case is the fraction by which the Andreev reflection process and the spin flip process at the interface get suppressed.
We can find out the angle integrated values of the above quantities by integrating over the angle of incidence. We defined them as,
\bea
G_{c\sigma}(E)=\frac{1}{G_N}\int_{\theta_c} d\theta cos\theta G_{c\sigma}(E,\theta)\\
G_{s\sigma}(E)=\frac{1}{G_N}\int_{\theta_c} d\theta cos\theta G_{s\sigma}(E,\theta)
\eea
where $G_N$ is the tunneling charge conductance from normal metal to normal metal and $\theta_c$ is the critical angle of incidence. We define two critical angles of incidence here. One is $\theta_c$ which comes from Equ (13) as $\theta_c=sin^{-1}(\frac{k_{\downarrow}^e}{k_{\uparrow}^e})$. For an incident angle $\theta>\theta_c$ there is no Andreev reflection process . There is another critical angle which we denoted by $\theta_{cs}$ and defined as $\theta_{cs}=sin^{-1}(\frac{k_{F+}}{k_{\uparrow}^e})$. For an angle of incidence greater than this angle there will be no transmitted particle. This restrictions on $\theta$ comes from the situation when the incoming electron is with up-spin. It is also noticeable that similar restriction does not arise for a down-spin incoming particle. In fact we can integrate the tunneling conductance for down spin particle over the entire angle of incidence, but since the up-spin particles impose some restrictions on the angle of incidence we have to follow that restriction in case of down-spin also. So the actual critical angle will be the smaller one between $\theta_c$ and $\theta_{cs}$. To determine this the relative magnitude of $\alpha$ and $E_{ex}$ is important. It can be easily shown that for $\alpha < E_{ex}$, $\theta_c<\theta_{cs}$. So the actual critical angle should then be $\theta_c$. In the reverse case $\theta_{cs}$ would define the appropriate limit of integration. 

Finally we define the total charge and spin conductance as,
\bea
G_c(E)=\Sigma_\sigma P_\sigma G_{c\sigma}(E)\\
G_s(E)=\Sigma_\sigma \sigma P_\sigma G_{s\sigma}(E)
\eea
where $\sigma$ is positive for up-spin and negative for down-spin incoming electrons. 

\maketitle
\section{result and discussion}
\begin{figure}
\includegraphics[height=8cm,width=16cm,angle=0]{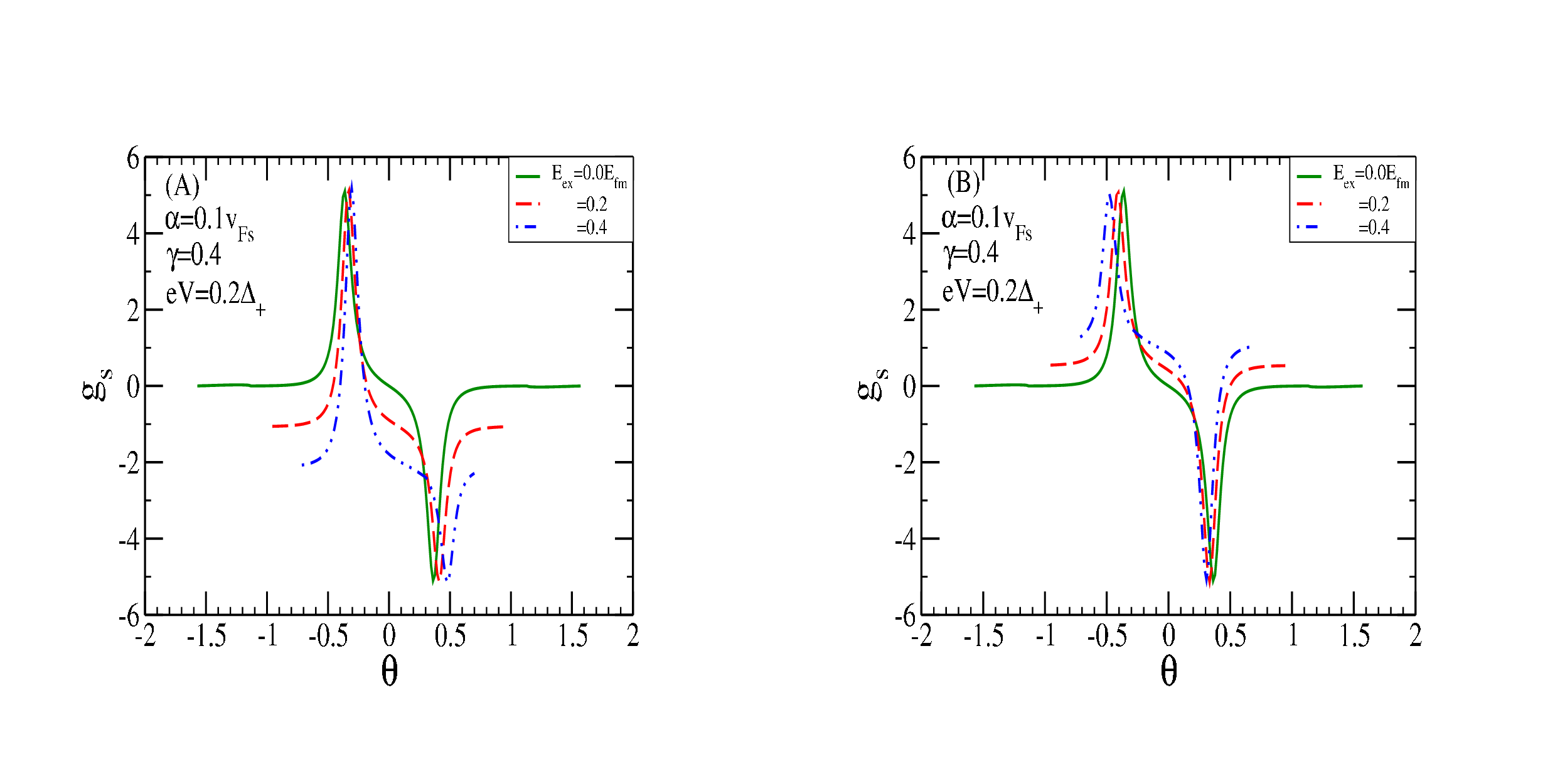}
\caption{(Color online) The variation of charge conductance $g_c$ and spin conductance $g_s$ (normalized with $G_N$) with incident angle $\theta$ for different values of exchange energy $E_{ex}$.The parameters $Z=5$,$\gamma=0.4$,$\alpha/v_{Fs} = 0.1$,$eV=0.2\Delta_+$ are chosen.The exchange energy are  0.0 (black solid line), 0.2 (red short-dashed line), 0.4 (green dashed-dot line). All in the unit of $E_{fm}$. 
In (A) the population in the up-spin sub-band is greater than the down-spin sub-band and in (B) the population in the down-spin sub-band is greater than the up-spin sub-band.
}
\label{fig:cond_theta}
\end{figure}

\section*{(A)Role of Andreev reflection in the tunneling process}
In the Andreev reflection process the energy as well as spin is conserved. When a normal metal is coupled to a s-wave superconductor then an electron with up-spin after Andreev reflection returns as a down-spin hole. As a result a Cooper pair is formed with charge $2e$ and spin zero ie. in singlet state. The presence of exchange energy in ferromagnetic metals gives rise to two different spin sub-bands with different population. The noncentrosymmetric superconductor can sustain both singlet as well triplet order parameters due to the presence of Rashba spin-orbit coupling and two bands are of different helicities. As a result when these two materials are coupled together then on the ferromagnetic side the incident electron with spin $\sigma$ coming from the $\sigma$ sub-band may be Andreev reflected as a hole in the $\overline{\sigma}=-\sigma$ sub-band with spin $\overline{\sigma}=-\sigma$ or as a hole in the $\sigma$ sub-band with spin $\sigma$. In the former process there is a tendency of formation of singlet Cooper pairs with opposite spin but in the later process triplet Cooper pairs with spin ($\overline{\sigma},\overline{\sigma}$) is formed to conserve the spin in the Andreev process. Although instead of propagating as separate entity they ultimately form quasiparticles which is a combination of both of them. So in this process some spin current is transported due to the edge states present in the noncentrosymmetric superconductor. Since in our sample the spin-up band is given some preferential population over the spin-down band so the triplet pair with both spin down will be produced preferentially. As a result a negative spin current will flow as is seen from the asymmetry of the Fig.~\ref{fig:cond_theta} which is opposite to the charge current. This spin current can be reversed by creating a greater population in the down-spin band. Although in other figures we have plotted the absolute value of this negative spin current so they appears positive.

\section*{(B)Discussions of the numerical results}
We have numerically solved equations 11,12 to get various coefficients and later they were numerically integrated to find out the values of the various angle integrated quantities. In Fig.~\ref{fig:cond_theta} we have shown the variation of both spin and charge conductances (normalized with $G_N$) with the variation of the angle of incidence $\theta$. Since they are the values of the angle resolved conductances so we denote them by $g_c$ and $g_s$. The presence of the exchange interaction is incorporated in such a way that it creates a preferential population to the spin-up sub-band. Also there is an preferential probability for the incident up-spins. This extra probability factor in ferromagnetic metal makes an $\theta$ dependent asymmetry in the charge as well as in the spin conductances and breaks the symmetry $g_c(\phi)\neq g_c(-\phi)$ (for angle resolved charge conductance) and $g_s(\phi)\neq -g_s(-\phi)$ (for angle resolved spin conductance) . As a result a spin current appears along with the conventional charge current across the junction due to the presence of the edge states in the NCS superconductor. This spin conductance arising due to the presence of the edge states and without the application of the magnetic field also modulates with the variation of the bias voltage across the junction. With the increase of the exchange energy the probability for up-spin incidence increases and the spin conductance looses the symmetry more and more as is seen in Fig.~\ref{fig:cond_theta}. We have shown in these diagrams the variation for the case when exchange energy is absent. In this case the variation is odd with respect to $\theta$ and the net spin current is zero when integrated over the angle of incidence. We have also shown the result(figure (B)) when the population in the down-spin sub-band is greater than the up-spin sub-band.

\begin{figure}
\centerline{\epsfysize=15cm\epsfbox{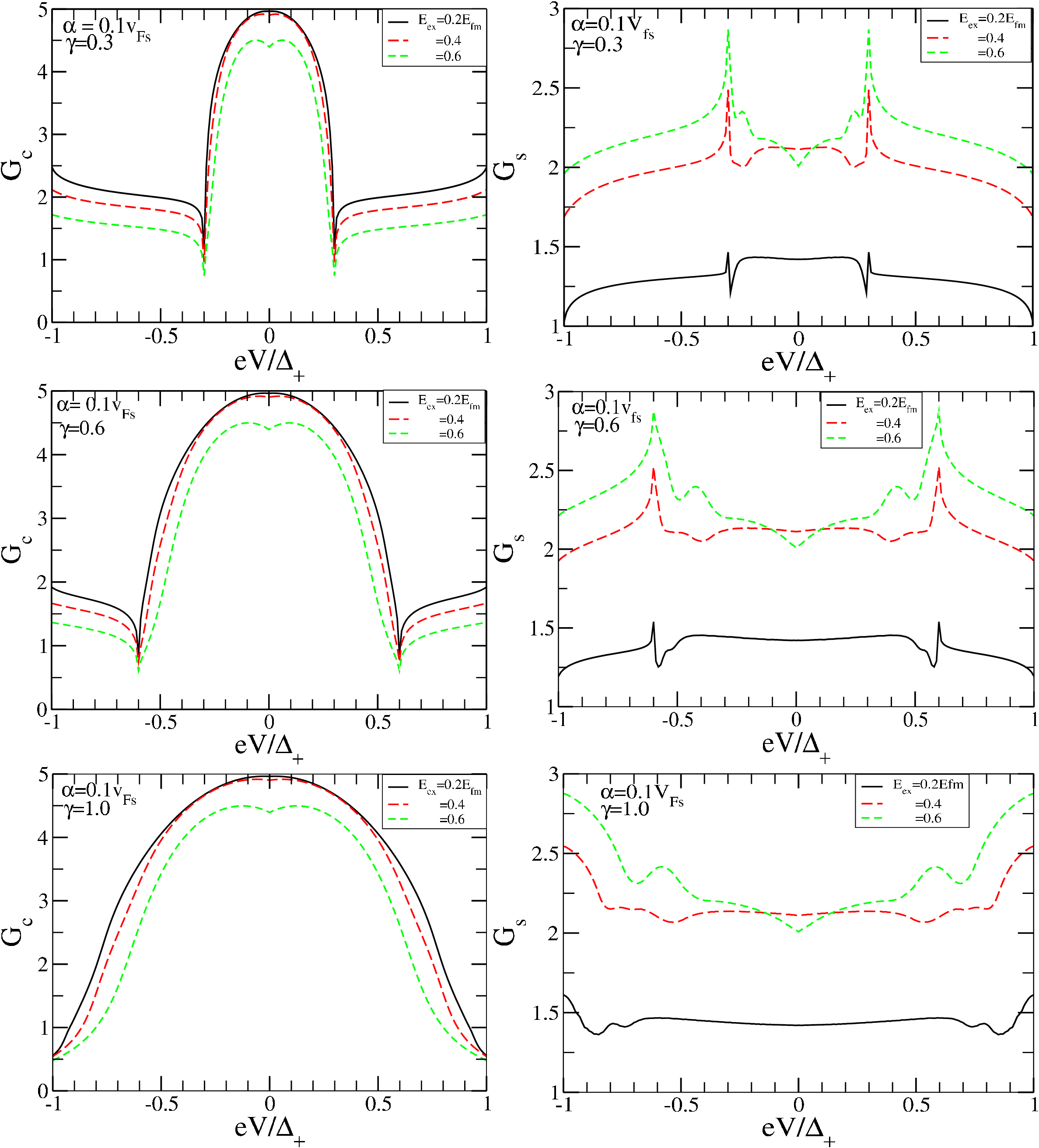}}
\caption{(Color online) The variation of charge conductance (left panel) and spin conductance (right panel) with bias energy $eV$ for different values of exchange energy $E_{ex}$. We have also shown the variation for different $\gamma$. The parameters $Z=3$ and $\alpha/v_F = 0.1$ are chosen. The exchange energy are  0.2 (black solid line), 0.4 (red long-dashed line), 0.6 (green short-dashed line). All in the unit of $E_{fm}$.
}
\label{fig:cond_exchange}
\end{figure}

\begin{figure}
\includegraphics[height=8cm,width=16cm,angle=0]{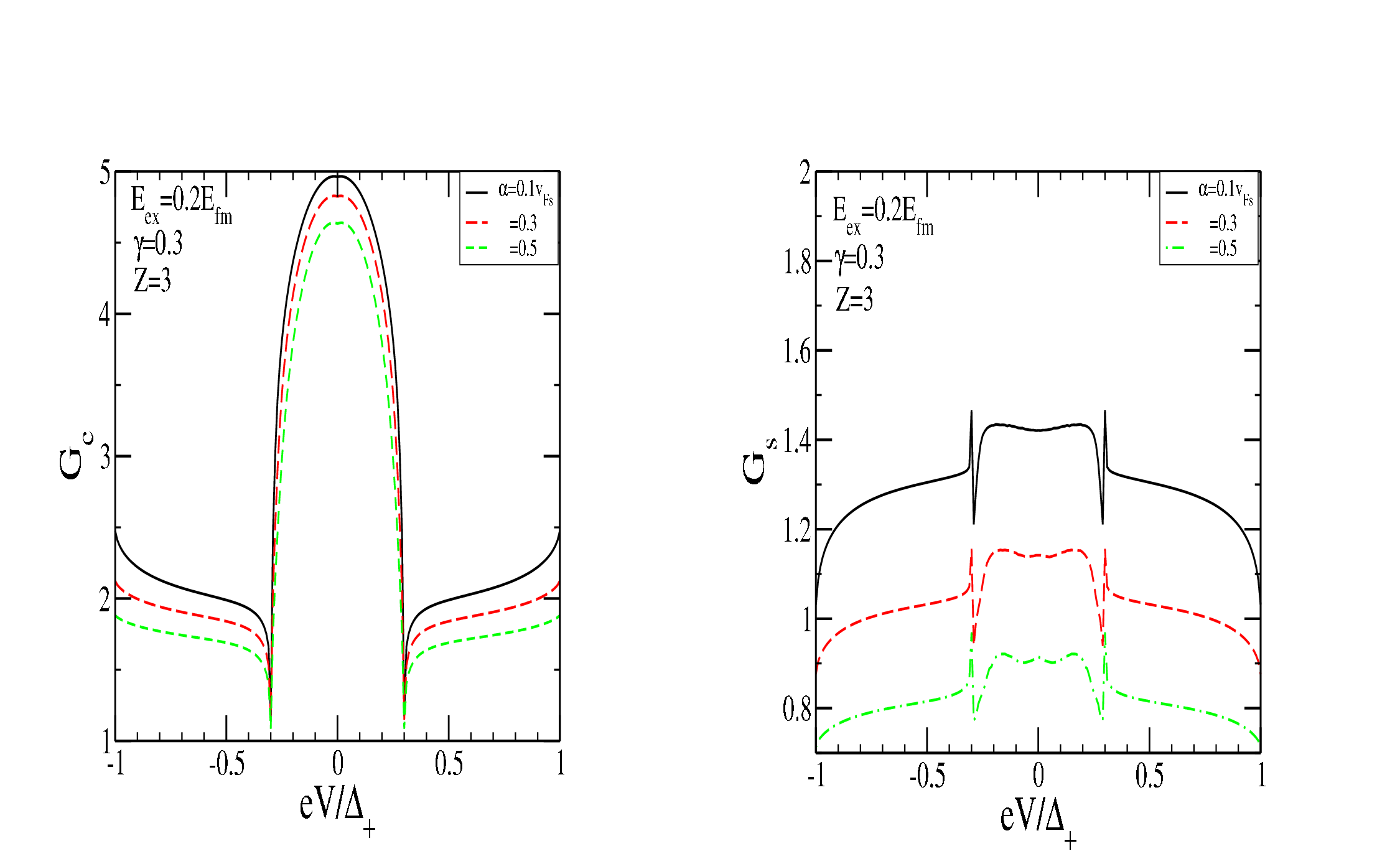}
\caption{(Color online) The variation of charge conductance $G_c$ and spin conductance $G_s$ with bias energy $eV$ for different values of Rashba parameter $\alpha/v_{Fs}$.The parameters $Z=3$,$\gamma=0.3$ and $E_{ex} = 0.2 E_{fm}$ are chosen.
}
\label{fig:cond_alpha}
\end{figure}

\begin{figure}
\includegraphics[height=12cm,width=12cm,angle=270]{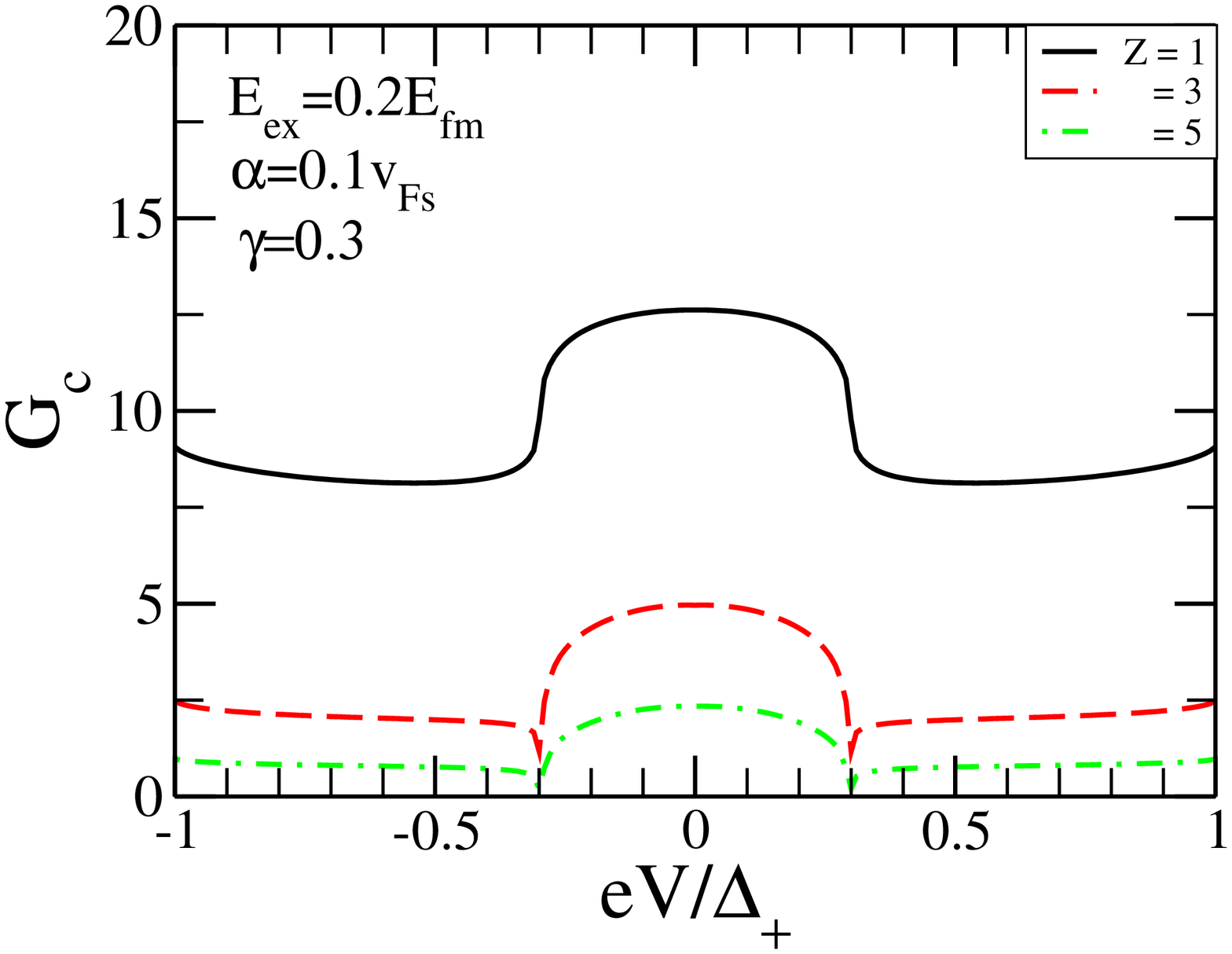}
\caption{(Color online) The variation of charge conductance with bias energy $eV$ for different values of barrier strength $Z$.
}
\label{fig:cond_Z}
\end{figure}

In Fig.~\ref{fig:cond_exchange} we have plotted the variation of angle integrated charge and spin conductances along with the bias voltage, normalized by the normal metal/normal metal tunneling conductance $G_N$. The structure has almost the same dome shaped feature as in the case of normal metal/NCS superconductor\cite{Mukherjee}. Here in each plot we have shown the variation of the charge and spin conductances for different values of the exchange energy. As expected the charge conductance decreases along with the increase of the exchange energy $E_{ex}$. The reason for this is that with the increase of the exchange energy the energy difference between the two opposite spin bands increases and as a result the probability of both normal as well as Andreev reflection decreases and hence there is a decrease in the charge conductance. Also with the increase of the exchange energy the critical angle of incidence $\theta_c$ decreases along with the critical angle inside the NCS superconductor $\theta_{cs}$ .However the spin conductance has somewhat opposite trends. Initially with the increase of the exchange energy as the energy difference between two opposite spin sub-bands increases the probability that an electron with a certain spin returns with the opposite hole after the Andreev reflection is reduced. Also with the increase of the exchange energy the probability that the incoming particle is with spin up increases. This seriously affects the spin conductance but have little effect on charge conductance. As a result the formation of Cooper pairs with the parallel spin triplet pairing becomes more viable compared with the singlet pairing and this gives rise to an increased spin current along with the increase of the exchange energy. This spin current though rapidly falls down when the exchange energy is increased more after a certain value.

We have also shown the feature of the charge as well as spin conductance for different ratios of the $\gamma=\frac{\Delta_-}{\Delta_+}$. For each $\gamma$ a dip appears in the conductance diagram exactly at the position where the bias voltage $eV$ equals $\gamma$. In fact this feature is universal and always appears whenever we will plot charge and spin conductances with the variation of $eV$ for any value of $\gamma$. One interesting feature of the figures for $\gamma\neq1$ is that the variation of the charge conductance for $eV>\gamma$ changes more with the increases of the exchange energy compared to that within the dome shaped region.It happens due to the fact that quasi-particles with  $eV>\gamma$ in the NCS superconducting region effectively see one band, the bound states of the other band simply becomes inaccessible. Within the dome shaped region however both the bands take active part in the conduction process and as a result the variation of the heights of the peaks of the charge conductance for small variation of the exchange energy becomes negligibly small.

In Fig.~\ref{fig:cond_alpha} we have shown the variation of the charge and spin conductances for different values of the Rashba parameter $\alpha$ keeping the other parameters fixed. This figure is also interesting. This figure shows that with the increase of $\alpha$ both charge and the spin conductance decrease. The reason is obvious. With the increase of $\alpha$ the critical angle $\theta_{cs}$ decreases. We would like to mention here that for $\alpha>E_{ex}$ the angle $\theta_c$ exceeds $\theta_{cs}$. As a result we have to perform the integration between$\pm \theta_{cs}$. So there is a competition between the Rashba parameter $\alpha$ and the exchange energy $E_{ex}$. When $\alpha<E_{ex}$ then $\theta_c$ is the critical angle but whenever we have the reverse situation i.e. $\alpha>E_{ex}$ then $\theta_{cs}$ becomes the critical angle for integration. Finally in Fig.~\ref{fig:cond_Z} we have shown the variation of the charge conductance with the variation of the barrier height. The conductance decreases with the increase of the barrier height and the behavior is as usual. 

\section{summary}

We summarize by saying that the helical edge states \cite{Nagaosa} exist in a noncentrosymmetric superconductor provided the triplet-pair amplitude is larger than the singlet-pair amplitude, i.e., when $0 < \gamma \leq 1$. We have studied the effect of these edge states on the tunneling charge and spin conductance between a ferromagnetic metal and a NCS superconductor. The presence of the exchange energy breaks $g_c(\phi)\neq g_c(-\phi)$ symmetry (for angle resolved charge conductance) and $g_s(\phi)\neq -g_s(-\phi)$ (for angle resolved spin conductance) by creating an probability imbalance between the number of up-spin and down-spin incident electrons and as a result there exist a tunneling spin current along with the usual charge current due to the presence of the edge states in NCS superconductor. Not only that, this spin current can be tuned by tuning the bias voltage across the junction and it changes shape with the change in the exchange energy in the ferromagnetic metal. The direction of this spin current is reversed by creating a reverse population density in the spin sub-bands. We have shown the variation of both the conductances with the different values of the ratios between the two order parameters which we call $\gamma$. The variation of the charge as well spin conductances with the change in the Rashba parameter $\alpha$ and the barrier height $Z$ are also presented. From these plots we can conclude that there is a competition between the Rashba parameter $\alpha$ and the exchange energy $\E_{ex}$ which manifests itself through the presence of two critical angles $\theta_c$ and $\theta_{cs}$ and the variation of charge and spin conductances . The plot of the variation of conductance with Z shows no unusual behavior.

\section*{Acknowledgment}
I gratefully acknowledge S.S. Mandal for the helpful discussions I made with him and the time and effort he gave for critically reviewing the manuscript and making many important suggestions.

\end{document}